# TRANSFORMITY: THE DEPENDENCE OF THE LAWS OF PHYSICS ON HIGHER-DIMENSIONAL COORDINATE TRANSFORMATIONS

Paul S. Wesson

Dept. of Physics and Astronomy, University of Waterloo, Waterloo, Ontario N2L 3G1, Canada

Abstract: In unified field theories with more than four dimensions, the form of the equations of physics in spacetime depends in general on the choice of coordinates in higher dimensions. The reason is that the group of coordinate transformations in (say) five dimensions is broader than in spacetime. This kind of gauge dependence is illustrated by two examples: a cosmological model in general relativity and a matter wave in quantum theory. Surprisingly, both are equivalent by coordinate transformations to flat featureless five-dimensional space. This kind of *transformity* is of fundamental significance for the philosophy of physics.

Email: psw.papers@yahoo.ca

# TRANSFORMITY: THE DEPENDENCE OF THE LAWS OF PHYSICS ON HIGHER-DIMENSIONAL COORDINATE TRANSFORMATIONS

1. <u>Introduction</u>

It is common to regard the laws of physics as in some sense immutable. The equations which embody the laws are widely considered to have forms that enjoy more permanence than other manifestations of human activity, such as the rules of justice or the edicts of religion. Even when certain gifted individuals show the way to more accurate laws, as from Newton to Einstein, the adjustment is incremental and not wholesale. This agrees with the widely-held view that the laws of science are objective and lie outside the realm of human interference. It will therefore probably come as a revelation to the practising physicist, or the informed layman, that this reverence for the laws of physics is misplaced.

It is widely believed that the best route to a theory that unifies the interactions of particles (quantum mechanics) with gravity (general relativity) is via extra dimensions. Currently, two such accounts are in vogue, namely membrane theory and space-time-matter theory (hereafter referred to as M theory and STM theory). These theories have similar mathematical structures, and both employ a five-dimensional abstract 'space' that embeds four-dimensional spacetime. Both are in agreement with available observational data. They achieve this by allowing the extra coordinate to play a significant role, contrary to the old 5D Kaluza-Klein theory [1]. Indeed, the flexibility provided by the extra coordinate gives 5D theory the conceptual elasticity necessary to stretch from microphysics to macrophysics. However, it appears to have slipped the attention of many physicists that the presence of an extra coordinate also has an unexpected consequence. In technical terms, a coordinate transformation which includes the fifth coordinate will in general change the form of the equations in the embedded *four* dimensions of spacetime. Some gauge transformations of this type can have a drastic effect, as will be seen below, effectively changing a cosmology into an elementary particle. In literary terms, it is as if a change in the language were to change the meaning of a sentence. If the world really does have extra dimensions, the basic equations for the physics of spacetime with which we have become familiar cannot be regarded as fixed.

Alarm and despondency might be the response among pedestrian physicists, perhaps followed by a rejection of the concept and with it the fifth dimension. However, the non-covariant behaviour of the 4D equations in 5D theory has been appreciated by a few discerning people, and regarded with equanimity. Einstein understood the situation, and in work with Bergmann went on to state: "We ascribe physical reality to the fifth dimension" [2]. Einstein's contemporary, Sir Arthur Eddington, developed a comprehensive philosophy of physics, in which certain subjective elements occur as a necessary consequence of how humans see the world [3]. A modern version of Eddington's philosophy, in which 5D physics figures prominently, is due to Wesson [4]. His account has much in common, as regards the treatment of the equations of physics, with the book of Hawking and Mlodinow [5], though the latter authors do not base their argument for a grand design on extra dimensions. As regards the effects of 5D coordinate transformations on 4D physics, the subject has been treated explicitly in two recent books [6]. It has also been examined by Ponce de Leon, who has shown in considerable detail how a simple solution in 5D can lead to several different but acceptable physical models in 4D [7]. The purpose of the present account is to bring this topic to a wider audience. While the issue is at base mathematical, it will become clear that there are broader implications of the dependence of the laws of physics on higher-dimensional coordinate transformations, which in short can be termed *transformity*.

In theories which involve relativity (which is essentially all of those in modern physics), the concept of transformity is related to that of covariance. The latter is the invariance of the 4D equations under changes of the labels for time and ordinary 3D space. Technically, transformity involves the breakdown of 4D covariance under coordinate transformations in $N$D where $N \geq 5$. However, this should actually be voiced in a positive way, since if appropriate transformations can be found, it is possible to generate new solutions to a set of field equations from a known one. This procedure is not restricted to finding 4D solutions from a 5D one. Einstein's field equations, for example, are commonly regarded as describing gravitational phenomena in a curved 4D spacetime. But those equations are not restricted to four dimensions, and have been extended to higher ones, up to versions of string theory with $N = 26$. In Riemannian geometry, as used for Einstein's theory of general relativity, it is always possible to embed an $N$D theory in an ($N$+1) D manifold. This is guaranteed by a local embedding theorem which was outlined in a book by Campbell, subsequently proved in detail in a Ph.D. thesis by Magaard, and then applied to modern theories of the Kaluza-Klein type [8]. Such 5D theories describe the gravitational and electromagnetic fields, with the equations of Einstein and Maxwell, plus a scalar field which is believed to concern the masses of particles and is frequently couched as an equation of the Klein-Gordon type. Depending on the application, the 5D formalism gives rise to the aforementioned STM theory [6] and M theory [9]. The former uses the fifth dimension to give a geometrical origin for matter, while the latter uses it to explain the masses of particles. As a theory of fields, 5D relativity is classical in nature. But the quantum analog involves gravitons with spin 2, photons with spin 1 and scalar particles with spin 0. The last may be related to the Higgs mechanism by which particles acquire masses, currently under study at the Large Hadron Collider. It becomes apparent that 5D relativity has considerable scope, and it accordingly affects many areas of physics [10]. It has, in particular, been employed to derive 5D models of the universe based on general relativity [11-15] and 5D wave-mechanical models of particles based on quantum theory [16-20]. Since it is inherent to 5D relativity, it also becomes apparent that *transformity* has considerable scope.

To illustrate the efficacy of transformity, two examples will be given in the next section. One concerns cosmology and the other concerns quantum theory. The working is mathematical, but the results are given a more qualitative interpretation in the last section.

2. <u>Two Examples of Transformity</u>

In order to match the nomenclature of 4D general relativity, the coordinates of a point in the 5D manifold are labelled $x^0 = t$ for time, $x^{123} = r\theta\phi$ or *xyz* for ordinary 3D space, and $x^4 = l$ for the extra length. (This is the usage in STM theory, while M theory commonly uses $x^4 = y$, a practise avoided here to forestall confusion.) Then the interval between two nearby points is given in terms of the metric tensor by $dS^2 = g_{AB} dx^A dx^B$ ($A, B = 0,123,4$ with summation over repeated indices). This contains the conventional 4D interval $ds^2 = g_{\alpha\beta} dx^\alpha dx^\beta$ ($\alpha, \beta = 0,123$). But because $g_{AB} = g_{AB}(x^\gamma, l)$ then generally $g_{\alpha\beta} = g_{\alpha\beta}(x^\gamma, l)$ and the 4D potentials may depend on the extra coordinate. For the purpose of the two examples to follow, it is not really necessary to assign the fifth coordinate $x^4 = l$ any special kind of physical meaning. However, those readers who wish to have a physical identification may like to note that there is a special class of metrics known as canonical, where it is assumed that the extra coordinate measures the rest mass *m* of a test particle [6, 15]. There are actually two ways of doing this, suggested by the fundamental constants for gravitation and quantum physics. Namely $l = Gm/c^2$ and $l = h/mc$, the Schwarzschild radius and the Compton wavelength. Here *G* is the gravitational constant, *h* is Planck's constant of action and *c* is the speed of light. To streamline the algebra, though, these constants can be absorbed via a suitable choice of units which renders their magnitudes equal to unity.

The field equations for 5D relativity are frequently taken to be the analogs of the 4D ones of general relativity as applied to the solar system. In terms of the Ricci tensor, they read

$$R_{AB} = 0 \quad (A, B = 0, 1, 2, 3, 4) \ . \tag{1}$$

These have 15 independent components, and numerous solutions of them are known [6]. It is now general knowledge that the 5D equations contain the 4D Einstein equations, by virtue of Campbell's embedding theorem [8]. Einstein's equations, in terms of the tensor named after him and the energy-momentum tensor, read

$$G_{\alpha\beta} + \Lambda g_{\alpha\beta} = 8\pi T_{\alpha\beta} \ . \tag{2}$$

The cosmological constant $\Lambda$ is included explicitly here because it will prove important in the second example below. It should be noted that the matter terms which appear on the right-hand side of (2) can, if so desired, be thought of as arising from the extra terms in (1) due to the fifth dimension. This is a corollary of Campbell's theorem, and is the reason why STM theory is sometimes referred to as induced-matter theory.

Flat space in 5D is simply an extension of Minkowski space in 4D, with interval

$$dS^2 = dT^2 - d\Sigma^2 \pm dL^2 \quad . \tag{3}$$

Here $d\Sigma^2$ is shorthand for the separation in 3D, so $d\Sigma^2 = dR^2 + R^2(d\theta^2 + \sin^2\theta d\phi^2)$ in spherical polar coordinates or $d\Sigma^2 = (dx^2 + dy^2 + dz^2)$ in Cartesian coordinates. The last term in (3) can have either sign, depending on whether the extra coordinate is spacelike or timelike. (It does not have the nature of a time, so there is no problem with closed timelike paths which lead to paradoxes with causality.) For the 5D canonical metric mentioned above, it is known that the sign choice in the metric is connected with the sign of the cosmological constant: $\Lambda > 0$ for a spacelike extra dimension, and $\Lambda < 0$ for a timelike extra dimension.

Transformity, as defined before, implies that a given situation in 5D can lead to different physical situations in 4D. This sounds blasphemous when judged from the viewpoint of classical general relativity. But that it can happen when the theory is widened by extra dimensions may be appreciated by considering the groups of coordinate transformations involved in 5D and 4D:

$$x^A \to \bar{x}^A(x^B) \quad , \qquad x^\alpha \to \bar{x}^\alpha(x^\beta) \quad . \tag{4}$$

These are not equivalent, the former being broader than the latter ($A = 0-5$, $\alpha = 0-4$). A little thought shows that, provided coordinate transformations involve the extra coordinate in a meaningful way, a single algebraic form in 5D can change to different forms in 4D. And insofar as we ascribe different physics to different algebraic models, a given 5D form can yield different kinds of 4D physics. Consequences of transformity occur throughout the last 20 years' work on unconstrained 5D relativity [6, 7, 11, 15]. It is hoped that the present account will serve to focus attention on the underlying cause, which is the discrepancy in the transformation groups (4). Arguably, the best way to illustrate the effects of transformity is to show how it is possible to go from the flat 5D space (3) to curved 4D spaces, which interpreted using general relativity describe situations with significant physics. That is, in loose language, to go from a featureless situation to ones with real physics.

Cosmology provides the first example. Consider the flat metric (3) with coordinates $T$, $R$, $\theta$, $\phi$, $L$ when the angles $\theta$, $\phi$ are held fixed but the other measures are changed to $t$, $r$, $l$ where

$$T(t,r,l) = \frac{\alpha}{2}\left[\left(1+\frac{r^2}{\alpha^2}\right)t^{1/\alpha}l^{1/(1-\alpha)} - \frac{t^{(2\alpha-1)/\alpha}l^{(1-2\alpha)/(1-\alpha)}}{(1-2\alpha)}\right]$$

$$R(t,r,l) = rt^{1/\alpha}l^{1/(1-\alpha)}$$

$$L(t,r,l) = \frac{\alpha}{2}\left[\left(1-\frac{r^2}{\alpha^2}\right)t^{1/\alpha}l^{1/(1-\alpha)} + \frac{t^{(2\alpha-1)/\alpha}l^{(1-2\alpha)/(1-\alpha)}}{(1-2\alpha)}\right] \quad . \tag{5}$$

Here $\alpha$ is a dimensionless constant whose significance will soon become clear. The extra dimension is taken to be spacelike in (3), which by some tedious algebra can be shown to become

$$dS^2 = l^2 dt^2 - t^{2/\alpha}l^{2/(1-\alpha)}\left(dr^2 + r^2 d\Omega^2\right) - \frac{\alpha^2 t^2}{(1-\alpha)^2}dl^2 \quad . \tag{6}$$

This metric is recognized as one of a class found originally by Ponce de Leon, who looked for 5D solutions to the field equations (1) that reduce on the hypersurfaces $l=$ constants to the standard Friedmann-Robertson-Walker ones of general relativity [11]. In fact, the 4D part of (6) describes FRW models with flat space sections and scale-factors $S(t) \sim t^{1/\alpha}$ for the expansion rate. (The angular part of the metric is written in shorthand using $d\Omega^2 \equiv d\theta^2 + \sin^2\theta d\phi^2$.) The choice $\alpha = 3/2$ gives back the Einstein-deSitter model with $S(t) \sim t^{2/3}$, which is the simplest realistic cosmology in general relativity. The interpretation of (6) as FRW models is confirmed by working out its properties of matter using Einstein's equations (2) and the 4D part of (6). Assuming that the matter is a perfect fluid with density $\rho$ and pressure $p$, these are found to be specified by

$$8\pi\rho = \frac{3}{\alpha^2\tau^2}, \qquad 8\pi p = \frac{2\alpha - 3}{\alpha^2\tau^2}, \qquad (7)$$

where $\tau \equiv lt$ is the proper time. The equation of state is $p = (2\alpha/3 - 1)\rho$. For $\alpha = 3/2$, the density and pressure are $\rho = 1/6\pi\tau^2$ with $p = 0$, which is the standard dust model. For $\alpha = 2$, $\rho = 3/32\pi\tau^2 = 3p$, which is the standard radiation model. In summary, reasonable models for the universe are obtained by applying transformity to flat space.

Quantum mechanics provides the second example. The first working model for quantum phenomenon was provided by the wave mechanics of de Broglie, which was cast into the form of an equation applicable to the hydrogen atom by Schrodinger. The general, relativistic form of the Schrodinger equation is the Klein-Gordon equation. While existing since 1924, wave mechanics has undergone a resurgence in modern times, with new versions of the double-slit experiment and renewed interest in neutron interferometry [20]. Since 5D relativity includes a scalar field believed to be related to the masses of particles, a challenge for it is to make contact with wave mechanics, and especially with the Klein-Gordon equation (which can be regarded as a kind of field equation with particle rest mass as source). Assuming that de Broglie waves propagate through the vacuum, it is natural to consider the de Sitter solution of general relativity. In this, the energy density of the vacuum is measured by the cosmological constant $\Lambda$ [6]. The de Sitter solution in Einstein's theory is commonly given in textbooks in two forms, a cosmological (expanding) version and a local (static) version. These are related by a somewhat involved (4D) coordinate transformation, whose precise form need not be repeated here. But it was shown long ago by Robertson that the cosmological version of de Sitter space could be embedded by a coordinate transformation in 5D Minkowski space [13]. And more recently Rindler has given a thorough demonstration of the same thing for the static version, showing that it maps to a kind of sphere in flat 5D space [14]. If the goal is to use transformity to go between flat space and a de Broglie wave propagating in the (de Sitter) vacuum, the remaining step is to connect the vacuum metric to a wave metric. This has been done by Wesson [17]. The details are not really necessary, since the flatness of the wave metric in question can be shown quickly by computer, as can the fact that it satisfies the field equations (1). The metric is of canonical form, with interval

$$dS^2 = \frac{l^2}{L^2}\left\{c^2 dt^2 - \exp\left[\pm\frac{2i}{L}(ct+\alpha x)\right]dx^2 - \exp\left[\pm\frac{2i}{L}(ct+\beta y)\right]dy^2 \right.$$

$$\left. - \exp\left[\pm\frac{2i}{L}(ct+\gamma z)\right]dz^2\right\} + dl^2 \quad . \tag{8}$$

Hence the constant length $L$ is related to the cosmological constant by $\Lambda = -3/L^2$. There is a wave in 3D, whose frequency is $f = 1/L$ and whose wave-numbers in the three directions of ordinary space are $k_x = \alpha/L$, $k_y = \beta/L$ and $k_z = \gamma/L$. The dimensionless constants $\alpha, \beta, \gamma$ are arbitrary, so the speed of the wave in the x-direction (say) is $c/\alpha$ and can exceed the speed of light $c$ (which is here kept explicit). Other properties of the wave show that it is of de Broglie type, and obeys his relation $v_p v_g = c^2$ between the phase velocity and the group velocity [18]. In wave mechanics, the phase velocity refers to the oscillations in the vacuum, while the group velocity is identified with the speed of the associated particle. These complementary phenomena are, of course, the essence of wave-particle duality. Incidentally, the equations of motion which go with the metric (8) yield the Klein-Gordon equation [16]. Considering the complexity of the physics implied by the wave metric (8), it is remarkable that the latter is related to the metric (3) of flat space.

3. Discussion and Conclusion

Transformity may be regarded as a kind of hidden principle of N-dimensional field theory, and there are good reasons why it is only now emerging as a subject of discussion. It says that, given a 'big' space of dimension (N+1) that contains a 'little' space of dimension N, the group of coordinate transformations for the former is broader that for the latter, so in general a change of coordinates in the big space will change the appearance of the equations of physics in the observer's little space. This was illustrated in the foregoing section, where two examples were given showing how a flat space in 5D can be changed to curved spaces in 4D, as relevant to cosmology and wave mechanics.

These cases fit naturally into the modern versions of 5D relativity known as space-time-matter theory and membrane theory. It is important to realize that in both of these theories, the 4D part of the 5D metric depends in general on the extra coordinate $x^4 = l$. This is quite different from the old Kaluza theory, where the so-called "cylinder condition" ruled out dependency of the metric coefficients on $l$ [1]. The modern theories also eschew the condition introduced by Klein known as "compactification", where the extra dimension is assumed to be rolled up to an unobservably small size. The richness of the algebra of unconstrained 5D relativity leads to situations where a given metric in 5D can correspond to different metrics in 4D, and thereby to different physics. The examples given above show this, though it should be pointed out that transformity is not restricted to 5D.

In a theory with $N$ dimensions, the field equations determine the components of the metric tensor $g_{AB}$, which is symmetric. Regarding $g_{AB}$ as an $N \times N$ array, there are $N^2$ components, but they are not all independent. Along the diagonal there are $N$ elements, so in each of the symmetric off-diagonal sectors there are $(N^2 - N)/2$ elements. This plus the number along the diagonal shows there are $(N^2 - N)/2 + N = N(N+1)/2$ independent components of $g_{AB}$. These are exactly determined by the field equations $R_{AB} = 0$, which has the same number of independent components. Both STM and M theory have $N = 5$, so there are 15 physically relevant components of the field equations. More complicated theories use $N > 5$. Increasing the dimensionality $N$ does not necessarily lead to improved understanding, because the physical meaning of the higher-dimensional coordinates becomes less clear. The field equations also become harder to solve as $N$ increases. However, in any theory of this type, the coordinates are arbitrary as they are in Einstein's 4D theory (covariance). For any $N$, there are $N$ degrees of freedom available to choose the coordinates in a way which renders the field equations easier to solve. For $N = 5$, four of these degrees of freedom are frequently used to remove the electromagnetic potentials ($g_{4\alpha}$), as was done for the metrics (6) and (8). For the latter case, the remaining degree of coordinate freedom was used to flatten the scalar potential ($g_{44}$), giving the canonical metric. Interestingly, the availability of $N$ degrees of coordinate freedom means that an $N$D metric can always be put into the form of an $(N-1)D$ one, plus an extra piece which is flat and therefore physically innocuous. This is another consequence of Campbell's theorem, which applies to any value of $N$ [8]. While special values of $N$ may have certain appeal because of the group structure, transformity implies that there is no 'magic' value of $N$.

Flatness is a special property of certain solutions to the $N$D field equations, which some workers find of special interest. However, it may not always be apparent. For example, the cosmological and wave-type metrics given above do not *look* flat, and this property is only revealed by fairly complicated coordinate transformations. It should also be emphasised that there are many solutions of the 5D field equations (1) which have physically-reasonable forms but are not flat [6]. That is, they have a vanishing Ricci tensor but a non-vanishing Riemann tensor ($R_{AB} = 0$, $R_{ABC}{}^D \neq 0$; the second has a greater number of independent components than the first). The best-known class of 5D solutions of this type are the 3D spherically-symmetric objects known as solitons. The 4D part of the 5D soliton metric is a generalization of the standard Schwarzschild metric of general relativity, and both have singularities at the centre of ordinary 3D space. It was demonstrated many years ago by Tangherlini and others that the Schwarzschild solution of Einstein's equations can only be embedded in a flat space if $N \geq 6$. To embed *any* solution of Einstein's equations in a flat space the latter must have $N \geq 10$, a property basic to the approach to unification known as supersymmetry.

In 5D, a preferred way to solve the field equations is to presume the existence of a gauge. This is commonly a function of the extra coordinate ($x^4$), which is multiplied onto a 4D metric that depends on the coordinates of spacetime ($x^\gamma$). The canonical metric of STM theory and the warp metric of M theory both involve gauges. An advantage of the gauge approach is that it separates the dependency of the physics on $x^4$ and $x^\gamma$, making for an easier interpretation. However, not all solutions of the field equations can be put into gauge form. And even if there is a well-defined gauge, it is not clear if the 4-space experienced by an observer is the whole of the 4D part of the metric or only its post-gauge $x^\gamma$-dependent part. (This problem of interpretation also exists in 4D scale-invariant theories, where a scalar function multiplies a 4D spacetime and the two choices are referred to as the Jordan and Einstein frames.) It should also be pointed out that a change to the gauge function, even a small or trivial one, can significantly alter the physics experienced by an observer confined to the 4D $x^\gamma$-dependent subspace. This happens, for example, in the canonical metric where a shift $(l \to l_0)$ in the extra coordinate causes a divergence to appear in the purely 4D cosmological 'constant' [6, 16]. This constitutes a special kind of transformity, and is one of the things that need further study. A possible application of gauge-like transformity concerns the interpretation of quantum mechanics due to Stueckelberg and Feynman [19]. They argued that, in order to correctly calculate the probability of a particle moving between one location and another in 4D spacetime, it is necessary to take into consideration *all* conceivable paths, not only the classically-preferred or shortest one. This *sum-over-paths* interpretation of quantum mechanics draws support from several directions, and has been adopted into cosmology by Hoyle and Narlikar, Hawking and others [6]. But where does it originate? A plausible explanation is that the many gauges of 5D manifest themselves as the sum-over-paths of a particle in 4D.

It has been seen that transformity can affect both cosmology and particle physics. Solutions of the field equations (1) relevant to those subjects are given by metrics (6) and (8). These look very different; but they are in fact equivalent via coordinate transformations to the metric (3) of flat 5D space. This is typical of transformity, which arises essentially because the group of 5D coordinate transformations is broader than the group of 4D transformations, so a change involving the fifth coordinate affects the 4D physics. An implication is that 5D physics concerns not only the search for solutions of the field equations, but also a search for appropriate coordinate systems. This may at first appear to be a drawback of 5D physics. But on second thought, such an opinion is recognized as coming from a viewpoint restricted to 4D, where the arbitrariness of coordinates is frequently regarded as a nuisance. Instead, transformity should be regarded as an integral part of the quest for unification. It applies in any unrestricted $N$D theory, and is in fact an inevitable consequence of the algebra of such theories. From the mathematical side, transformity should be regarded as an algebraic technique, akin to others in field theory such as covariance. From the physical side, it fits with the views of Eddington and others [3, 4] who argue that to a certain extent the laws by which the world is described depend on how humans perceive it.

Acknowledgements

Thanks for comments go to J.M. Overduin and other members of the Space-Time-Matter group (http://5Dstm.org).